\documentclass{amsart}
\usepackage{amssymb, amsmath, epsfig}
\usepackage{amscd}

\newtheorem{remark}{\it Remark}
\newtheorem{theorem}{Theorem}

\newtheorem{corollary}{Corollary}

\newtheorem{conjecture}{Conjecture}
\newtheorem{lemma}{Lemma}

\newtheorem{theoremA}{Theorem A}

\newcommand{\R}{\mathbb{R}}

\newcommand{\ddim}{\mathrm{ddim\;}}
\newcommand{\dind}{\mathrm{dind\;}}
\newcommand{\corank}{\mathrm{corank\;}}
\newcommand{\so}{\mathfrak{so}}
\newcommand{\Sym}{\mathrm{Sym}}
\newcommand{\gl}{\mathfrak{gl}}

\newcommand{\ad}{\mathrm{ad}}
\DeclareMathOperator{\Ad}{\mathrm{Ad}}
\DeclareMathOperator{\Span}{\mathrm{span\,}}
\DeclareMathOperator{\diag}{\mathrm{diag}}

\DeclareMathOperator{\rank}{\mathrm{rank}}
 \DeclareMathOperator{\pr}{\mathrm{pr}}
\DeclareMathOperator{\tr}{\mathrm{tr}}

\title{Singular Manakov Flows and Geodesic Flows on Homogeneous Spaces}

\author{Vladimir Dragovi\' c, Borislav Gaji\' c and Bo\v zidar
Jovanovi\'c}

\begin{document}
\baselineskip=14pt \maketitle
\begin{abstract}
We prove complete integrability of the Manakov-type
$SO(n)$-inva\-riant geodesic flows on homogeneous spaces
$SO(n)/SO(k_1)\times\dots\times SO(k_r)$, for any choice of
$k_1,\dots,k_r$, $k_1+\dots+k_r\le n$. In particular, a new proof
of the integrability of a Manakov symmetric rigid body motion
around a fixed point is presented. Also, the proof of
integrability of the $SO(n)$-invariant Einstein metrics on
$SO(k_1+k_2+k_3)/SO(k_1)\times SO(k_2)\times SO(k_3)$ and on the
Stiefel manifolds $V(n,k)=SO(n)/SO(k)$ is given.
\end{abstract}

\section{Introduction}

It was Frahm who gave the first four-dimensional generalization of
the Euler top in the second half of XIX century, \cite{Fr}.
Unfortunately, his paper was forgotten for more than a century. A
modern history of higher-dimensional generalizations of the Euler
top has more than thirty years after the paper of Manakov in 1976
\cite{Ma}. Manakov used Dubrovin's theory of matrix Lax opeartors
(see \cite {Dub, Du}) to prove that the solutions of
Frahm-Manakov's top can be given in terms of theta-functions.
Although the subject has had intensive development since then,
there are still few questions which in our opinion  deserve
additional treatment.

\subsection{Liouville Integrability}
Let $(M,\{\cdot,\cdot\})$ be a Poisson manifold. The equations
\begin{equation}
\dot f=\{f,H\}, \quad f\in C^\infty (M)
\label{HAM}
\end{equation}
are called {\it Hamiltonian equations} with the {\it Hamiltonian function}
$H$. A function $f$ is an integral of the Hamiltonian system (constant
along trajectories of (\ref{HAM})) if and only if it
commutes with $H$: $\{f,H\}=0$.

One of the central problems in Hamiltonian dynamics is whether the
equations (\ref{HAM}) are completely integrable or not. The
equations (\ref{HAM}) are  {\it completely integrable} or {\it
Liouville integrable} if there are $l=\frac12(\dim M
+\corank\{\cdot,\cdot\})$ Poisson-commuting smooth integrals
$f_1,\dots,f_l$ whose differentials are independent in an open
dense subset of $M$. The set of integrals $\mathcal
F=\{f_1,\dots,f_l\}$ is called a {\it  complete involutive  set}
of functions on $M$. To distinguish the situation from the case of
non-commutative integrability, the last set will be called {\it
commutative} as well.

If the system is completely integrable, by the Liouville-Arnold
theorem there is an implicitly given set of coordinates in which
the system trivializes.  Moreover, from the  Liouville-Arnold
theorem \cite{Ar} follows that regular compact connected invariant
submanifolds given by integrals $\mathcal F$ are Lagrangian tori
within appropriate symplectic leaves of the Poisson bracket
$\{\cdot,\cdot\}$ and the dynamics over the invariant tori is
quasi-periodic.

\subsection{Noncommutative Integrability}

 Let
$(M,\{\cdot,\cdot\})$ be a Poisson manifold, $\Lambda$ be the
associated bivector field on $M$
$$
\{f,g\}(x)=\Lambda_x(df(x),dg(x))
$$
and let $\mathcal F$ be a Poisson subalgebra of $C^\infty(M)$ (or
a collection of functions closed under the Poisson bracket).

Consider the linear spaces
\begin{equation}
 F_x=\{ df(x) \, \vert\, f\in\mathcal F\}\subset T_x^* M
\label{NABLAF}
\end{equation}
and suppose that we can find  $l$ functionally independent
functions $f_1,\dots,f_l \in \mathcal F$ whose differentials span
$F_x$ almost everywhere on $M$ and that the corank of the matrix
$\{f_i,f_j\}$ is equal to some constant $k$, i.e., $\dim
\ker\Lambda_x\vert_{F_x}=k$.

The numbers $l$ and $k$ are called {\it differential dimension}
and {\it differential index} of $\mathcal F$ and they are denoted
by $\ddim\mathcal F$ and $\dind\mathcal F$, respectively.  The set
$\mathcal F$ is called {\it complete} if (see \cite{N, MF2, BJ2}):
$$
\ddim{\mathcal F}+\dind{\mathcal F}=\dim M+\corank\{\cdot,\cdot\}.
$$

The Hamiltonian system (\ref{HAM}) is {\it completely integrable
in the noncommutative sense} if it possesses a complete set of
first integrals $\mathcal F$. Then (under compactness condition)
$M$ is almost everywhere foliated by $(\dind{\mathcal
F}-\corank\{\cdot,\cdot\}$)-dimensional invariant tori. As in the
Liouville-Arnold theorem, the Hamiltonian flow restricted to
regular invariant tori is quasi-periodic (see  Nekhoroshev
\cite{N} and Mishchenko and Fomenko \cite{MF2}.

\subsection{Mishchenko--Fomenko Conjecture}
Let $\mathcal F$ be any Poisson closed subset of $C^\infty(M)$,
then a subset ${\mathcal F}^0 \subset \mathcal F$ is a {\it
complete subset} if $ \ddim {\mathcal F}^0 + \dind {\mathcal F}^0
= \ddim \mathcal F + \dind \mathcal F. $ In particular, a
commutative subset ${\mathcal F}^0 \subset \mathcal F$ is complete
if $ \ddim {\mathcal F}^0 = \frac12\left(\ddim \mathcal F + \dind
\mathcal F\right). $

Mishchenko and Fomenko stated the conjecture {\it that
non-commutative integrable systems are integrable in the usual
commutative sense by means of integrals that belong to the same
functional class as the original non-commutative integrals}. In
other words, if $\mathcal F$ is a complete set, then we can always
construct a complete commutative set ${\mathcal F}^0\subset
\mathcal F$.

Let us mention two cases in which the Mishchenko-Fomenko
conjecture has been proved. The  finite-dimensional version of the
conjecture is recently  proved by Sadetov \cite{Sa} (see also
\cite{Bo2, ViYa}): {\it for every finite-dimensional Lie algebra
$\mathfrak g$ one can find a complete commutative set of
polynomials on the dual space $\mathfrak g^*$ with respect to the
usual Lie-Poisson bracket}. The second case where the conjecture
was proved is  $C^\infty$--smooth case for infinite-dimensional
algebras (see \cite{BJ2}).

Consider the homogeneous spaces $G/H$ of a compact Lie group $G$.
Fix some $\Ad_G$-invariant positive definite scalar product
$\langle \cdot ,\cdot \rangle$ on the Lie algebra $\mathfrak
g=Lie(G)$. Let $\mathfrak h=Lie(H)$ and let $\mathfrak g=\mathfrak
h \oplus \mathfrak v$ be the orthogonal decomposition with respect
to $\langle \cdot,\cdot\rangle$. The scalar product $\langle
\cdot,\cdot \rangle $ induces
 a {\it normal} $G$-invariant metric on $G/H$ via
$ (\cdot,\cdot)_0=\langle\cdot,\cdot\rangle\vert_\mathfrak v, $
where $\mathfrak v$ is identified with the tangent space at the
class of the identity element. If $G$ is semisimple and
$\langle\cdot,\cdot\rangle$ is negative Killing form, the normal
metric is called {\it standard} \cite{Be}. The geodesic flow of
the normal metric is completely integrable in the non-commutative
sense  by means of integrals polynomial in momenta \cite{BJ1, BJ3}
and the Mishchenko-Fomenko conjecture can be reduced to the
following ones:

\begin{conjecture} {\rm(\cite{BJ3})}
{\rm For every homogeneous space $G/H$ of a compact Lie group $G$
there exist a complete commutative set of $\Ad_H$-invariant
polynomials on $\mathfrak v$. Here the Poisson structure is
defined by (\ref{THIMM}).}
\end{conjecture}

For example if $(G,H)$ is a spherical pair, the set of
$\Ad_H$-invariant polynomials is already commutative. In many
examples, such as Stiefel manifolds, flag manifolds, orbits of the
adjoint actions, complete commutative algebras are obtained (see
\cite{BJ1, BJ3, MP}), but the general problem is still unsolved.

Note that  solving the problem of commutative integrability of
geodesic flows of normal metrics would allow  to construct new
examples of $G$-invariant metrics on homogeneous spaces $G/H$ with
integrable geodesic flows as well.

\subsection{The Manakov Flows}
The Euler equations of a left-invariant geodesic flow on $SO(n)$
have the form
\begin{equation}
\dot M=[M,\Omega], \quad \Omega=\mathfrak A(M)
\label{Euler}
\end{equation}
where $\Omega\in \so(n)$ is the angular velocity, $M\in \so(n)^*\cong \so(n)$ angular momentum
and $\mathfrak I=\mathfrak A^{-1}$ the positive definite operator which defines the left invariant metric (see \cite{Ar}).
Here we identify $\so(n)$ and $\so(n)^*$
by means of the scalar product proportional to the Killing form
\begin{equation}
\langle X,Y\rangle=-\frac12\tr(XY),
\label{SP}
\end{equation}
$X,Y \in \so(n)$.
The Euler equations (\ref{Euler}) are Hamiltonian
with respect to the Lie-Poisson bracket
\begin{equation}
\{f,g\}(M)=-\langle M, [\nabla f(M),\nabla g(M)]\rangle, \qquad M\in\so(n)
\label{Lie-Poisson}
\end{equation}
with the Hamiltonian function $H=\frac12\langle M,\mathfrak A
(M)\rangle$. The invariant polynomials $\tr(M^{2k})$,
$k=1,\dots,\rank\so(n)$ are central functions, determining the
regular symplectic leaves (adjoint orbits) of the Lie-Poisson
brackets (\ref{Lie-Poisson}). Thus we need half of the dimension
of the generic adjoint orbit ($\frac12(\dim\so(n)-\rank\so(n))$)
additional independent commuting integrals for the integrability
of Euler equations (\ref{Euler}). For a generic operator
$\mathfrak A$ and $n \ge 4$ the system is not integrable.

Manakov  found the Lax representation
with rational parameter
$\lambda$ (see \cite{Ma}):
\begin{equation}
\dot L(\lambda)=[L(\lambda),U(\lambda)], \quad L(\lambda)=M+\lambda A, \quad U(\lambda)=\Omega+\lambda B,
\label{Lax}
\end{equation}
provided $M$ and $\Omega$ are connected by
\begin{equation}
[M,B]=[\Omega,A],
\label{Manakov}
\end{equation}
where $A$ and $B$ are diagonal matrices $A=\diag(a_1,\dots,a_n)$,
$B=\diag(b_1,\dots,b_n)$.

In the case the eigenvalues of
$A$ and $B$ are distinct, we have
\begin{equation}
\mathfrak A=\ad_A^{-1}\circ \ad_B=\ad_B \circ\ad_A^{-1} \quad
\Longleftrightarrow \quad
\Omega_{ij}=\frac{b_i-b_j}{a_i-a_j}M_{ij}, \label{operator}
\end{equation}
where $M_{ij}=\langle M, E_i\wedge E_j\rangle$. Here
$\ad_A(M)=[A,M]$ and $\ad_B(M)=[B,M]$ are considered as linear
transformations from $\so(n)$ to the zero-diagonal subspace of the
space of symmetric matrices $\Sym(n)$. They are invertible since
the eigenvalues of $A$ and $B$ are distinct. Note that we take $A$
and $B$ such that $\mathfrak A$ is positive definite. Formally, we
can take singular $B$ (i.e., $B$ with some equal eigenvalues), but
then $\mathfrak A$ is not invertible and represents the operator
which determines the left-invariant sub-Riemannian metric on
$SO(n)$.

The left invariant metric given by the operator (\ref{operator})
is usually called the {\it Manakov metric}. In this case, Manakov
proved that the solutions of the Euler equations (\ref{Euler}) are
expressible in terms of $\theta$-functions by using the
algebro-geometric integration procedure developed by Dubrovin in
\cite{Dub, Du} (see \cite{Ma}).


The explicit verification that
integrals
arising from the Lax representation
\begin{equation}
\mathcal L=\{\tr(M+\lambda A)^k\, \vert\, k=1,2,\dots,n,\, \lambda\in \mathbb{R}\},
\label{integrals}
\end{equation}
 form a
complete Poisson-commutative set on $\so(n)$ was given by
Mishchenko and Fomenko in \cite{MF1} in the case when the
eigenvalues of $A$ are distinct (see also Bolsinov \cite{Bo}).
Furthermore, the system is algebraically completely integrable.
Conversely, if a diagonal metrics $\Omega_{ij}=\mathfrak A_{ij}
M_{ij}$ with distinct $\mathfrak A_{ij}$ has algebraically
completely integrable geodesic flow then $\mathfrak A$ has the
form (\ref{operator}) for certain $A$ and $B$ (see \cite{Ha}). An
another Lax pair of the system can be found in \cite{Fe}.

\subsection{Singular Manakov Flows}
We shall describe operators $\mathfrak A$ satisfying the condition
(\ref{Manakov}) when the eigenvalues of $A$ and $B$ are not all
distinct. Suppose
\begin{eqnarray}
&&a_1=\dots=a_{k_1}=\alpha_1, \; \dots, \;a_{n+1-k_r}=\dots=a_n=\alpha_r,\nonumber \\
&& b_1=\dots=b_{k_1}=\beta_1, \; \dots \;
b_{n+1-k_r}=\dots=b_n=\beta_r, \label{beta} \\
&& k_1+k_2+\dots+k_r=n, \quad \alpha_i \ne \alpha_j, \quad \beta_i\ne \beta_j,\quad  i,j=1,\dots,r. \nonumber
\end{eqnarray}

Let
\begin{equation}\so(n)=\so(n)_A\oplus\mathfrak v=\so(k_1)\oplus
\so(k_2)\oplus\dots\oplus \so(k_r)\oplus \mathfrak v \label{v}
\end{equation}
be the orthogonal decomposition, where $ \so(n)_A=\{X\in \so(n)\,
\vert\, [X,A]=0\}. $ By $M_{\so(n)_A}$ and $M_\mathfrak v$ we
denote the projections of $M\in\so(n)$ with respect to (\ref{v}).

Further, let $\mathfrak B: \so(n)_A\to \so(n)_A$ be an arbitrary
positive definite operator. We take $A$ and $B$ such that the
sectional operator $\mathfrak A: \so(n)\to \so(n)$ defined via
\begin{equation}
\mathfrak A(M_{\mathfrak
v}+M_{\so(n)_A})=\ad_A^{-1}\ad_B(M_{\mathfrak v})+\mathfrak
B(M_{\so(n)_A}), \label{A}
\end{equation}
is positive definite. Now $\ad_A$ and $\ad_B$ are considered as
invertible linear transformations from $\mathfrak v$ to
$[A,\mathfrak v]\subset \Sym(n)$.

For the given $\mathfrak A$ we have $[\Omega,A]=[\Omega_\mathfrak
v,A]=[M_\mathfrak v,B]=[M,B], $ and the Manakov condition
(\ref{Manakov}) holds. It can be proved that $ [M_\mathfrak
v,\ad_A^{-1}\ad_B (M_\mathfrak v)]_{\so(n)_A}=0. $ Therefore, the
system (\ref{Euler}) takes the form
\begin{eqnarray}
&&\dot M_{\so(n)_A}=[M_{\so(n)_A},\mathfrak B (M_{\so(n)_A})], \label{s0} \\
&&\dot M_\mathfrak v=[M_{\so(n)_A},\ad_A^{-1}\ad_B (M_\mathfrak
v)]+ [M_\mathfrak v,\mathfrak B (M_{\so(n)_A})].\label{s1}
\end{eqnarray}

If $k_i \ge 4$ for some $i=1,\dots,r$, the equations (\ref{s0}) (and therefore
 the system (\ref{s0}), (\ref{s1})) are not integrable for a
generic $\mathfrak B$. On the other hand, since (\ref{Manakov})
holds, the system has Lax representation (\ref{Lax}). But the
integrals arising from the Lax representation do not provide
complete integrability.

We refer to (\ref{s0}), (\ref{s1}) as a {\it singular Manakov
flow}.

\subsection{Symmetric Rigid Bodies}
Consider the case $A=B^2$ and $\mathfrak A=\ad_{B^2}^{-1}\ad_B$.
Then the angular momentum and velocity are related by $M=\mathfrak
I(\Omega)=\ad_B^{-1}\ad_{B^2}(\Omega)=B\Omega+\Omega B$, i.e.,
\begin{equation}
\Omega_{ij}=\frac{1}{b_i+b_j}M_{ij}
\label{operator2}\end{equation} and the Euler equations
(\ref{Euler}), in coordinates $M_{ij}$, read
\begin{equation}
\dot M_{ij}=\sum_{k=1}^n \frac{b_i-b_j}{(b_k+b_i)(b_k+b_j)} M_{ik}M_{kj}.
\label{Euler3}
\end{equation}

The equations (\ref{Euler3}) describe the motion of a free
$n$-dimensional rigid body with a mass tensor $B$ and inertia
tensor $\mathfrak I$ around a fixed point \cite{FeKo}.

Now, in addition, suppose  that (\ref{beta}) holds (the case of a
$SO(k_1)\times SO(k_2)\times \dots\times SO(k_r)$--{\it symmetric
rigid body}). The operator $\mathfrak A$ given by
(\ref{operator2}) is well defined on the whole $\mathfrak{so}(n)$
and the restriction of $\mathfrak A$ to $\mathfrak{so}(k_i)$ is
the multiplication by $1/2\beta_i$. Thus, the system (\ref{s0}) is
trivial and we have the Noether conservation law
$M_{\so(n)_A}=const.$

Let us denote the set of linear functions on $\so(n)_A$ by
$\mathcal S$. These additional integrals provide the integrability
of the system. The complete integrability of the system is proved
by Bolsinov by using the pencil of Lie algebras on $\so(n)$ (see
the last paragraph of Section 2).

\subsection{Outline of the Paper} In Section 2 we prove that
Manakov integrals $\mathcal L$ together with  Noether integrals
$\mathcal S$ form a complete noncommutative set of polynomials on
$\so(n)$, giving  a new proof for the integrability of symmetric
rigid body motion (\ref{Euler3}). We also prove that Manakov
integrals induce a complete commutative set within
$SO(n)$-invariant polynomials on the cotangent bundle of the
homogeneous space $ SO(n)/SO(k_1)\times\dots\times SO(k_r) $ in
Section 3. The complete $SO(n)$-invariant commutative sets were
known before only for certain choices of numbers $k_1,\dots,k_r$
(see \cite{BJ1, BJ3}). In particular, it is proved in Section 4
that the construction implies the integrability of the
$SO(n)$-invariant Einstein metrics on
$SO(k_1+k_2+k_3)/SO(k_1)\times SO(k_2)\times SO(k_3)$ and on the
Stiefel manifolds $V(n,k)$. These Einstein metrics have been
obtained in \cite {Jen, Nik, adn, ADN}.

\section{Integrability of a Symmetric Rigid Body Motion}

\subsection{Completeness of Manakov Integrals}
Since the algebra of linear functions $\mathcal S$ is not
commutative if some of $k_1,\dots,k_r$ are greater than $2$, the
natural framework in studying singular Manakov flows is
noncommutative integration. We start with an   equivalent
definition of the completeness. Let $\mathcal F$ be a collection
of functions closed under the Poisson bracket on the Poisson
manifold $M$. We say that $\mathcal F$ is {\it complete at $x$} if
the space $F_x$ given by (\ref{NABLAF}) is coisotropic:
\begin{equation}
F^\Lambda_x \subset F_x \,.
\label{coisotropic}
\end{equation}
Here $F^\Lambda_x$ is skew-orthogonal complement of $F_x$ with respect to $\Lambda$:
$$
F^\Lambda_x=\{\xi\in T^*_x M\, \vert \, \Lambda_x(F_x,\xi)=0\}.
$$

The set $\mathcal F$ is {\it complete } if it is complete at a
generic point $x\in M$. In this case
$\ddim \mathcal F=\dim F_x$ and $F^\Lambda_x=\ker\Lambda_x\vert_{F_x}$ implying
$\dind \mathcal F=\dim F^\Lambda_x$, for
a generic $x\in M$.

Note that one can consider Hamiltonian systems restricted to symplectic
leaves as well. Let $N\subset M$ be a symplectic leaf (regular or
singular). The set $\mathcal F$ is complete on the symplectic leaf $N$ at
$x\in N$ if
\begin{equation}
F^\Lambda_x\subset F_x+\ker\Lambda_x
\label{coisotropic*}
\end{equation}
and it is {\it complete on the symplectic leaf} $N$ if it
complete at a generic point $x\in N$.

As above, let $\mathcal S$ be the set of linear functions on $\so(n)_A$
and $\mathcal L$ be the Lax pair integrals (\ref{integrals}).

\begin{theorem}\label{1}
$\mathcal L+\mathcal S$ is a complete noncommutative set of
functions on $\so(n)$ with respect to the Lie-Poisson bracket
(\ref{Lie-Poisson}). \end{theorem}

\begin{corollary}
The symmetric rigid body system (\ref{Euler3}), (\ref{beta}) is
completely integrable in the noncommutative sense. Moreover,
suppose that the system (\ref{s0}) is completely integrable on
$\so(n)_A$ with a complete set of commuting integrals $\mathcal
S^0$. Then the singular Manakov flow (\ref{s0}), (\ref{s1}) is
also completely integrable with a complete commuting set of
integrals $\mathcal L+\mathcal S^0$.
\end{corollary}

\noindent{\it Proof of theorem \ref{1}.} Let
$L_M=\{\nabla_M\tr(M+\lambda A)^k\, \vert\, k=1,2,\dots,n,\,
\lambda\in \mathbb{R}\}. $ According to (\ref{coisotropic}),
$\mathcal L+\mathcal S$ is complete at $M$ if
\begin{equation}
(L_M+\so(n)_A)^\Lambda\subset L_M+\so(n)_A,
\label{L}\end{equation} where $\Lambda$ is the canonical
Lie-Poisson bivector on $\so(n)$:
\begin{equation}
\Lambda(\xi_1,\xi_2)\vert_M=-\langle M, [\xi_1,\xi_2]\rangle.
\label{*}
\end{equation}

Consider the Lie algebra $\gl(n)$ of $n\times n$ real matrixes
equipped with the scalar product (\ref{SP}). We
have the symmetric pair
orthogonal decomposition $\mathfrak{gl}(n)=\so(n)\oplus\Sym(n)$ on the skew-symmetric and symmetric
matrices:
$$[\mathfrak{so}(n),\mathrm{Sym}(n)]\subset \Sym(n), \quad [\Sym(n),\Sym(n)]\subset \so(n).$$
The scalar product $\langle\cdot,\cdot\rangle$ is positive
definite on $\so(n)$ while it is negative definite on $\Sym(n)$.

Let us identify $\gl(n)^*$ and $\gl(n)$ by means of
$\langle\cdot,\cdot\rangle$. On $\gl(n)$ we have the pair of
compatible Poisson bivectors (see Reyman \cite{R})
\begin{eqnarray}
&&\Lambda_1(\xi_1+\eta_1,\xi_2+\eta_2)\vert_{X}=
-\langle X, [\xi_1,\xi_2]+[\xi_1,\eta_2]+[\eta_1,\xi_2]\rangle,\nonumber\\
&&\Lambda_2(\xi_1+\eta_1,\xi_2+\eta_2)\vert_X=-\langle X+A,[\xi_1+\eta_1,\xi_2+\eta_2]\rangle,
\label{PL}\end{eqnarray}
where $X\in \gl(n)$, $\xi_1,\xi_2\in \so(n)$, $\eta_1,\eta_2\in \Sym(n)$.
In other words, the pencil
$$
\Pi=\{\Lambda_{\lambda_1,\lambda_2}\; \vert\;  \lambda_1,\lambda_2\in \mathbb{R}, \;\lambda_1^2+\lambda_2^2\ne 0\},
\qquad \Lambda_{\lambda_1,\lambda_2}=\lambda_1 \Lambda_1+\lambda_2 \Lambda_2
$$
consist of Poisson bivectors on $\gl(n)$.

The Poisson bivectors $\Lambda_{\lambda_1,\lambda_2}$, for
$\lambda_1+\lambda_2\ne 0$ and  $\lambda_2 \ne 0$, are isomorphic
to the canonical Lie-Poisson bivector (in particular, their corank
is equal to $n$). The union of their Casimir functions
\begin{equation}
\mathcal F=\{f_{\lambda,k}(X)=\tr(\lambda M+ P +\lambda^2 A)^k\,
\vert\, k=1,2,\dots,n,\, \lambda\in \mathbb{R}\}\label{central}
\end{equation}
where $X=M+P$, $M\in \so(n)$, $P\in \Sym(n)$,
is a commutative set with respect to the all brackets from the
pencil $\Pi$ \cite{R, Bo}. Also, the skew-orthogonal complement $F^\Lambda_X$ does not depend on
the choice $\Lambda\in\Pi$. As above, $F_X$ denotes the linear
subspace of $\gl(n)$ generated by the differentials of
functions from ${\mathcal F}$ at $X$.

We need to take all
objects complexified (see \cite{Bo}). The complexification of
$\gl(n)$, $\so(n)$, $\Sym(n)$, $\so(n)_A$, $\Pi$ are
$\gl(n,\mathbb{C})$, $\so(n,\mathbb{C})$, $\Sym(n,\mathbb{C})$,
$\so(n,\mathbb{C})_A\cong \so(k_1,\mathbb{C})\oplus \dots\oplus\so(k_r,\mathbb{C})$ and
$\Pi^{\mathbb{C}}=\{\Lambda_{\lambda_1,\lambda_2}=\lambda_1\Lambda_1+
\lambda_2\Lambda_2$,
$\lambda_1,\lambda_2\in \mathbb{C}, \;
\vert\lambda_1\vert^2+\vert\lambda_2\vert^2\ne 0\}$, respectively.
Here, we consider (\ref{PL}) as complex valued
skew-symmetric bilinear forms. The complexified scalar product is simply
given by (\ref{SP}), where now $X,Y \in \gl(n,\mathbb C)$.

At a generic point $X\in \gl(n)$, the only singular bivector in
$\Pi^\mathbb{C}$ with a rank smaller then $\dim \gl(n)-n$ is
$\Lambda_{-1,1}$. Moreover, the complex dimension of the linear space
\begin{equation*}
K_{-1,1}=\{\xi\in \ker \Lambda_{-1,1}(X)\,
\vert \, \Lambda_0(\xi,\ker \Lambda_{-1,1}(X))=0\}\label{K}
\end{equation*}
is equal to $n$. Here $\Lambda_0$ is any Poisson bivector from the pencil, nonproportional to
$\Lambda_{-1,1}$, say $\Lambda_0=\Lambda_{0,1}$.
Whence, it follows from Proposition 2.5 \cite{Bo} that
$$
(F^{\Lambda_1}_X)^{\mathbb{C}}=F^\mathbb{C}_X+\ker\Lambda_{-1,1}(X).
$$

Also, it can be proved that
\begin{equation}
F^\mathbb{C}_X+\ker\Lambda_{-1,1}(X)=F^\mathbb{C}_X+\so(n,\mathbb{C})_A\,.
\label{SUM}
\end{equation}

The above relations imply
\begin{equation}
(F_X+\so(n)_A)^{\Lambda_1}=F_X^{\Lambda_1} \cap
\so(n)_A^{\Lambda_1} \subset F_X+\so(n)_A \label{GL}
\end{equation}
and the set of functions $\mathcal F+\mathcal S$ is a complete
non-commutative set on $\gl(n)$ with respect to $\Lambda_1$ (theorem 1.5 \cite{Bo},
for the detail proofs of the above
statements, given for an arbitrary semi-simple symmetric pair, see
\cite{TF}, pages 234-237).

Now we want to verify the completeness of $\mathcal F+\mathcal S$
at the points $M\in \so(n)$. Note that in theorem 1.6 \cite{Bo}, a
similar problem have been studied but for regular $A$ and singular
points $M\in \so(M)$, in proving that Manakov integrals provide
complete commutative sets on singular adjoint orbits.

Since a  regular matrix $M\in \so(n)$ ($\dim \so(n)_M=\rank\so(n)=[n/2]$), considered as an element of
$\gl(n,\mathbb{C})$ is also regular ($\dim\gl(n,\mathbb{C})_M=n$),
it can be easily proved that the
only two singular brackets in $\Pi^\mathbb{C}$ are
$\Lambda_{-1,1}$ and $\Lambda_{1,0}$.

We have to estimate the complex dimensions of linear spaces
\begin{eqnarray}
&& K_{-1,1}=\{\xi\in \ker \Lambda_{-1,1}(M)\,
\vert \, \langle M+A, [\xi,\ker \Lambda_{-1,1}(M)]\rangle=0\}\label{K1}\\
&&
K_{1,0}=\{\xi\in\ker \Lambda_{1,0}(M)\, \vert\, \langle M+A, [\xi,\ker \Lambda_{1,0}(M)]\rangle=0\}.
\label{K2}\end{eqnarray}

As for $X\in \gl(n)$, repeating the arguments of \cite{TF}, pages 234-237, one
can prove that the dimension of (\ref{K1}) is $n$ for a generic $M\in \so(n)$.
Further
\begin{equation}
\ker\Lambda_{1,0}(M)=\ker\Lambda_1(M)=\so(n,\mathbb{C})_M+\Sym(n,\mathbb{C}),
\label{KER}
\end{equation}
where $\so(n,\mathbb{C})_M$ is the isotropy algebra of $M$ in $\so(n,\mathbb{C})_M$.

We shall prove below that $\dim_{\mathbb C} K_{1,0}$ is also equal to $n$ for a generic $M\in \so(n)$ (see Lemma 1).
Hence, according Proposition 2.5 \cite{Bo}, at a generic $M\in \so(n)$ we have
\begin{equation}
(F_M^{\Lambda_1})^\mathbb{C}=
F_M^\mathbb{C}+\ker\Lambda_{-1,1}(M)+\ker\Lambda_{1,0}(M)=F_M^\mathbb{C}
+\so(n,\mathbb{C})_A+\ker\Lambda_1(M).
\label{SKEW*}
\end{equation}
Similarly as in equation (\ref{GL}) we get
\begin{eqnarray*}
\left(F_M+\so(n)_A\right)^{\Lambda_1} &=&
(F_M+\so(n)_A+\ker\Lambda_1(M))\cap \so(n)_A^{\Lambda_1}\\ &\subset & F_M+\so(n)_A+\ker\Lambda_1(M)\,.
\end{eqnarray*}

Therefore the relation (\ref{coisotropic*}) holds for functions
$\mathcal F+\mathcal S$ and the bracket $\Lambda_1$, i.e., this is
a complete set on the symplectic leaf through $M$.

Notice that the symplectic leaves through $M\in \so(n)\subset
\gl(n)$ of the bracket $\Lambda_1$ are $SO(n)$-adjoint orbit in
$\so(n)$ and that the restriction of $\Lambda_1$ to $\so(n)$
coincides with the Lie-Poisson bracket (\ref{*}). Since the
restriction of the central functions (\ref{central}) to $\so(n)$
are Manakov integrals (\ref{integrals}), we  obtain (\ref{L}).
$\Box$

\begin{remark}{\rm From the  proof of Theorem \ref{1} follows that the
skew-orthogonal complement of $L_M$ within $\so(n)$ is  given by
\begin{equation}
L_M^\Lambda = L_M+\so(n)_A\,,
\label{SKEW}\end{equation}
for a generic $M\in\so(n)$.
}\end{remark}

\begin{lemma}
The complex dimension of the linear space (\ref{K2}) is equal to $n$ for a generic $M\in \so(n)$.
\end{lemma}

\noindent{\it Proof.}
For $\xi\in \ker \Lambda_1(M)$, let $\xi_1$ and $\xi_2$ be the projections to
$\so(n,\mathbb{C})_M$ and $\Sym(n,\mathbb{C})$, respectively.
Then
\begin{eqnarray*}
\langle M+A, [\xi,\ker \Lambda_1(M)]\rangle
&=& \langle \ker \Lambda_1(M) , [M+A,\xi_1+\xi_2]\rangle \\
&=&     \langle \ker \Lambda_1(M) , [M,\xi_2]+[A,\xi_1]+[A,\xi_2]\rangle        \\
&=& \langle \so(n,\mathbb{C})_M, [A,\xi_2] \rangle + \langle \Sym(n,\mathbb{C}), [M,\xi_2]+[A,\xi_1] \rangle
\end{eqnarray*}

Therefore $\xi=\xi_1+\xi_2\in\ker \Lambda_1(M)$ belongs to $K_{1,0}$ if and only if
\begin{equation}
[M,\xi_2]+[A,\xi_1]=0, \quad \pr_{\so(n,\mathbb{C})_M} [A,\xi_2]=0.
\label{xi}
\end{equation}

The dimension of the solutions of the system (\ref{xi}), for a regular $M\in \so(n)$, is $n$.
It can be directly calculated by taking the following anti-diagonal element:
\begin{eqnarray*}
&&M=m_1 E_1 \wedge E_n + m_2 E_2\wedge E_{n-1}+\dots+ m_k E_{k} \wedge E_{k+1}, \quad n=2k\\
&&M=m_1 E_1 \wedge E_n + m_2 E_2\wedge E_{n-1}+\dots+ m_k E_{k}
\wedge E_{k+2}, \quad n=2k+1,
\end{eqnarray*}
when
\begin{eqnarray*}
&& \so(n,\mathbb{C})_M=\Span_\mathbb{C} \{E_1 \wedge E_n, E_2\wedge E_{n-1},\dots,E_{k}\wedge E_{k+1}\}, \quad n=2k\\
&&\so(n,\mathbb{C})_M=\Span_\mathbb{C} \{E_1 \wedge E_n, E_2\wedge E_{n-1},\dots,E_{k}\wedge E_{k+2}\}, \quad n=2k+1.
\end{eqnarray*}
Here $m_1,\dots,m_k$ are generic distinct real numbers. For
example, if $n=2k$, then $\xi\in \ker \Lambda_1(M)$ satisfies
(\ref{xi}) if and only if it is of the form:
$$
\xi=\sum_{i=1}^n u_i E_i\otimes E_i+\sum_{j=1}^k v_j E_j \wedge E_{n+1-j},
$$
where parameters $u_i,v_j$ are determined from the linear system:
\begin{equation*}
 - m_j(u_j-u_{n+1-j})+(a_j-a_{n+1-j})v_j=0, \quad j=1,\dots,k.
\end{equation*}

Thus $\dim_\mathbb{C} K_{1,0}=n$.
$\Box$

\subsection{Pencil of Lie Algebras} Bolsinov has given  another
proof of the integrability of Manakov flows, related to the
existence of compatible Lie algebra brackets on $\so(n)$
\cite{Bo}. The first Lie bracket is standard one
$[M_1,M_2]=M_1M_2-M_2M_1$, while the second is
$$
[M_1,M_2]_A=M_1AM_2-M_2AM_1.
$$
Then $\Lambda$ and $\Lambda_A$ are compatible Poisson structures,
where $\Lambda$ is given by (\ref{*}) and
\begin{equation}
\Lambda_A(\xi_1,\xi_2)\vert_M=-\langle M, [\xi_1,\xi_2]_A \rangle.\label{**}
\end{equation}

Let $\Lambda_{\lambda_1,\lambda_2}=\lambda_1 \Lambda+\lambda_2 \Lambda_A$.
The central functions of the bracket $\Lambda_{\lambda,1}$ of maximal rank
($\lambda \ne -\alpha_1,\dots,-\alpha_r$) are
\begin{equation}
\mathcal J=\{\tr(M(\lambda\mathbb{I}+A)^{-1})^{2k}\, \vert\, k=1,2,\dots,[n/2],
\lambda \ne -\alpha_1,\dots,\alpha_r\}.
\label{J}\end{equation}

According to the general construction, these functions commute
with respect to  all Poisson brackets
$\Lambda_{\lambda_1,\lambda_2}$. The following theorem obtained by
Bolsinov can be found in \cite{TF}, pages 241-244:

\begin{theorem}\label{2} {\rm (Bolsinov)}
The set of functions $\mathcal J+\mathcal S$ is a complete set on
$\so(n)$ with respect to the Lie-Poisson bracket
(\ref{Lie-Poisson}).
\end{theorem}

The families (\ref{integrals}) and (\ref{J})  commute between each
other  (e.g., see \cite{P}). Therefore, since both sets $\mathcal
L+\mathcal S$ and $\mathcal J+\mathcal S$ are complete, they are
equivalent, i.e., they determine the same invariant toric
foliation of the phase space.\footnote{It was pointed out by  one
of the referees that the equivalence of the integrals
(\ref{integrals}) and (\ref{J}) can be derived directly, by using
the  matrix identity $\det(M(A+\alpha\mathbb{I})^{-1}+\beta
\mathbb{I})=\det(M+\beta A+\alpha\beta
\mathbb{I})\det(\alpha\mathbb{I}+ A)^{-1}$.}

\medskip

\section{Geodesic Flows on $SO(n)/SO(k_1)\times\dots\times SO(k_r)$}

\subsection{Geodesic Flows on Homogeneous Spaces}
Consider the homogeneous spaces $G/H$ of a compact Lie group $G$.
Let $\mathfrak g=\mathfrak h \oplus \mathfrak v$ be the orthogonal
decomposition and let $ds^2_0$ be the normal $G$--invariant metric
induced by some $\Ad_G$-invariant scalar product $\langle \cdot
,\cdot \rangle$ on the Lie algebra $\mathfrak g$.

Let $\mathcal F^G$ be the set of $G$ invariant functions,
polynomial in momenta and $\Phi:T^*(G/H)\to \mathfrak g^*$ be the
momentum mapping of the natural Hamiltonian $G$-action. From the
Noether theorem we have $\{\mathcal F^G,\Phi^*(\R[\mathfrak
g^*])\}=0$, where $\{\cdot,\cdot\}$ is the canonical Poisson
bracket on $T^*(G/H)$. The Hamiltonian of the normal metric
$ds^2_0$ is a central function of $\mathcal F^G$ so it commutes
both with the Noether functions $\Phi^*(\R[\mathfrak g^*])$ and
$G$-invariant functions $\mathcal F^G$. On the other side, the set
$\mathcal F^G+\Phi^*(\mathbb R[\mathfrak g^*])$ is complete,
implying the noncommutative integrability of the geodesic flow of
the normal metric \cite{BJ1, BJ2}.

The algebra $(\mathcal F^G, \{\cdot,\cdot\})$ can be naturally
identified with $(\R[\mathfrak v]^H, \{\cdot,\cdot\}_\mathfrak
v)$, where $\R[\mathfrak v]^H$ is the algebra of $\Ad_H$-invariant
polynomials on $\mathfrak v$ and (see Thimm \cite{Th}):
\begin{equation}
\{f,g\}_\mathfrak v(x)=-\langle x, [\nabla f(x),\nabla g(x)]
\rangle, \qquad  f,g \in \R[\mathfrak v]^H. \label{THIMM}
\end{equation}

Within the class of Noether integrals $\Phi^*(\R[\mathfrak g^*])$
one can always construct a complete commutative subset. Thus the
Mishchenko--Fomenko conjecture reduces to a construction of a
complete commutative subset of $\mathbb{R}[\mathfrak v]^{H}\cong
\mathcal F^G$. This leads to Conjecture 1 stated in the
Introduction.

A commutative set $\mathcal F \subset \mathbb{R}[\mathfrak v]^{H}$
is complete if
\begin{equation}
\ddim \mathcal F =\frac12\left(\ddim \mathbb{R}[\mathfrak
v]^{H}+\dind\mathbb{R}[\mathfrak v]^{H}\right)= \dim
\mathfrak v-\frac12 \dim {\mathcal O}_{G} (x),
\end{equation}
for a generic $x\in \mathfrak v$, where ${\mathcal O}_{G}(x)$
is the adjoint orbit of $G$ (see \cite{BJ1, BJ3}).

\subsection{Normal Geodesic Flows on
$SO(n)/SO(k_1)\times\dots\times SO(k_r)$} Let
$$
SO(n)_A=SO(k_1)\times\dots\times SO(k_r)\subset SO(n)
$$
be the isotropy group of $A$ within $SO(n)$ with respect to the
adjoint action. As above, consider the normal metric $ds^2_0$
defined by the scalar product (\ref{SP}) and identify
$SO(n)$--invariant polynomials on $T^*(SO(n)/SO(n)_A)$ with
$\mathbb{R}[\mathfrak v]^{SO(n)_A}$ ($\mathfrak v$ is defined by
(\ref{v})).

We shall use the following completeness criterium.
Consider the space $\mathfrak j_M \subset \mathfrak v$ spanned by gradients of
all polynomials in $\mathbb{R}[\mathfrak v]^{SO(n)_A}$. For a
generic point $M\in \mathfrak v$ we have
$$
\mathfrak j_M=([M,\so(n)_A]^\perp)\cap \mathfrak v=\{ \eta\in
\mathfrak v \, \vert\,  \langle \eta,[M,\so(n)_A]\rangle=0\}
=\{\eta\in \mathfrak v\, \vert\, [M,\eta]\subset \mathfrak v\}.
$$

The bracket (\ref{THIMM}) on
$\mathbb{R}[\mathfrak v]^{SO(n)_A}$ corresponds to the restriction
of the Lie-Poisson bivector (\ref{*}) to $\mathfrak j_M$. Denote this
restriction by $\bar\Lambda$. Then $\mathcal F\subset\mathbb{R}[\mathfrak v]^{SO(n)_A}\cong \mathcal F^{SO(n)}$ is a complete
commutative set if and only if
$$
F_M^{\bar\Lambda}=F_M,
$$
for a generic $M\in\mathfrak v$, where $ F_M=\Span \{\nabla_M
f(M)\, \vert \, f\in\mathcal F\}\subset \mathfrak j_M
$
and $F_M^{\bar\Lambda}$ is the skew-orthogonal complements of
$F_M$ with respect to $\bar\Lambda$ within $\mathfrak j_M$. Here,
for simplicity, the gradient operator with respect to the
restriction of $\langle\cdot,\cdot\rangle$ to $\mathfrak v$ is
also denoted by $\nabla$.

Since all polynomials in $\mathcal L$ commute with $\mathcal S$,
their restrictions to $\mathfrak v$
\begin{equation}
{\mathcal L}_{\mathfrak v}=\{\tr(M+\lambda A)^k\, \vert\, M\in
\mathfrak v, \, k=1,2,\dots,n,\, \lambda\in \mathbb{R}\},
\label{LV} \end{equation}
 form a commutative subset of
$\mathbb{R}[\mathfrak v]^{SO(n)_A}$ (see \cite{BJ1}).

Let $\Phi: T^*SO(n)/SO(n)_A\to \so(n)^*\cong \so(n)$ be the
momentum mapping of the natural $SO(n)$-Hamiltonian action on
$T^*SO(n)/SO(n)_A$ and let $\mathcal A$ be any commutative set of polynomial on
$\so(n)$ that is complete on adjoint orbits within the image $\Phi(T^*(SO(n)/SO(n)_A))$
(for example one can take Manakov integrals with regular $A$ \cite{Bo}).
Then $\Phi^*(\mathcal A)$ is a complete commutative  subset in $\Phi^*(\R[\so(n)])$ and we have:

\begin{theorem}\label{homogeneous}
{\rm (i)} ${\mathcal L}_\mathfrak v$ is a complete commutative
subset of $\mathbb{R}[\mathfrak v]^{SO(n)_A}$.

{\rm (ii)} The geodesic flow of the normal metric $ds^2_0$ is
Liouville integrable by means of polynomial integrals ${\mathcal
L}_\mathfrak v+\Phi^*(\mathcal A)$.
\end{theorem}

\begin{remark}{\rm
Note that, by using the {\it chains of subalgebras method}, the
construction of another complete commutative algebras of
$SO(n)$-invariant polynomials is performed for homogeneous spaces
$SO(n)/SO(k)$ and  $SO(n)/SO(k_1)\times SO(k_2)$ (see \cite{BJ1,
BJ3}). Also, by using the {\it generalized chains of subalgebras
method}, the Conjecture 1 is solved for a class of homogeneous
spaces $SO(n)/SO(n)_A$ \cite{BJ3}. The class, say $\mathcal C$, is
obtained by induction from $SO(n)/SO(k_1)\times SO(k_2)$, $k_1\le
k_2 \le [\frac{n+1}2]$ in the following way: suppose that
$SO(n_1)/SO(k_1)\times\dots\times SO(k_{r_1})$ and
$SO(n_2)/SO(l_{1})\times\dots\times SO(l_{r_2})$ ($n_1=n_2\pm
0,1$) belong in $\mathcal C$, then also
$SO(n_1+n_2)/SO(k_1)\times\dots\times SO(k_{r_1})\times
SO(l_{1})\times\dots\times SO(l_{r_2})$ belongs to $\mathcal C$ .
Note that, for example, the homogeneous spaces
$SO(n)/SO(k_1)\times\dots\times SO(k_r)$, where some of $k_i$ is
grater than $[\frac{n+1}2]$ do not belong to the family $\mathcal
C$.}\end{remark}

\noindent{\it Proof.} Without loss of generality, suppose
$$
k_1\le k_2\le k_3\dots \le k_r.
$$
If the condition
\begin{equation*}
k_r \le \left[\frac{n+1}2\right]
\label{regular}
\end{equation*}
is satisfied, then a generic element $M\in \mathfrak v$ is regular element of $\so(n)$
and relation (\ref{SKEW}) will holds. Then it easily follows that
\begin{equation}
\bar L_M^{\bar\Lambda}=\bar L_M, \qquad \bar L_M=\{\nabla_M f\,
\vert\, f\in{\mathcal L}_\mathfrak v\}\subset \mathfrak j_M,
\label{L2}
\end{equation}
for a generic $M\in\mathfrak v$. Hence ${\mathcal L}_\mathfrak v$
is complete.

Now, suppose
$$
k_r=\left[\frac{n+1}2\right]+l, \quad l>0.
$$

Let $n'=n-2l$, $k'_r=k_r-2l$, $ A'=\diag(a_1,a_2,\dots,a_{n'})$
and let
\begin{equation}
\so(n')=\so(n')_{A'}\oplus \mathfrak v'=\so(k_1)\oplus
\so(k_2)\oplus\dots\oplus \so(k'_r)\oplus \mathfrak v' \label{v'}
\end{equation}
be the orthogonal decomposition, where $\so(n')_{A'}$ is the
isotropy algebra of $A'$ within $\so(n')$.

Furthermore, we can consider Lie algebras $\so(n')$ and $\so(2l)$
embedded in $\so(n)$ as blocks:
$$
\begin{pmatrix}
\so(n') & 0 \\
0 & \so(2l)
\end{pmatrix}.
$$

Then the linear space $\mathfrak v'$ becomes a linear subspace of
$\mathfrak v$:
$$
\mathfrak v'=so(n')\cap \mathfrak v.
$$

Moreover, for an arbitrary $M\in\mathfrak v$ one can find a matrix
$K\in SO(n)_A$ such that $M'=\Ad_{K}(M)$ belongs to $\mathfrak
v'$. Indeed, consider $M$ and $K$
 of the form
$$
M=\begin{pmatrix}
M_{11} & M_{12} \\
-M_{12}^T & 0
\end{pmatrix}, \qquad
K=\begin{pmatrix}
I_{n-k_r} & 0 \\
0 & U
\end{pmatrix},
$$
where $M_{11}\in \so(n-k_r)$, $M_{12}$ is $(n-k_r)\times (k_r)$
matrix, $I_{n-k_r}$ is the identity $(n-k_r)\times (n-k_r)$ matrix
and $U \in SO(k_r)$. Then
$$
M'=KMK^{-1}=\begin{pmatrix}
M_{11} & M_{12}U^T \\
-UM_{12}^T & 0
\end{pmatrix}.
$$
Since $k_r-(n-k_r)$ is equal to $2l$ or $2l+1$, one can always
find $U$ such that the last $2l$ rows of $UM_{12}^T$, i.e., the
last $2l$ columns of $M_{12} U^T$ are equal to zero, which implies
that $M'$ belongs to $\mathfrak v'$.

Therefore, if the set ${\mathcal L}_\mathfrak v$ is complete at
the points of $\mathfrak v'$ then it will be complete on
$\mathfrak v$ as well.

The Lie algebras $\so(n')$ and $\so(n')_{A'}$ are centralizers
of $\so(2l)$ in $\so(n)$ and $\so(n)_A$, respectively.
Whence, from the above considerations, we can apply Theorem A1 (see Apendix)
to get
\begin{equation}
\mathfrak j'_{M'}=\{ \eta\in \mathfrak v'\, \vert \langle
\eta,[M',\so(n')_A]\rangle=0\}= \{ \eta\in \mathfrak v\, \vert
\langle \eta,[M',\so(n)_A]\rangle=0\}=\mathfrak j_{M'},
\label{mik}
\end{equation}
for a generic $M'\in \mathfrak v'$.

In particular, (\ref{mik}) implies that Poisson tensors $\bar\Lambda$ of
$\mathbb{R}[\mathfrak v]^{SO(n)_A}$ and $\bar\Lambda'$ of $\mathbb{R}[\mathfrak v']^{SO(n')_{A'}}$
coincides on a generic $M'\in\mathfrak v'\subset\mathfrak v$.
According to the first part of the proof, the set
of polynomials
$$
{\mathcal L}_{\mathfrak v'}=\{\tr(M'+\lambda A')^k\, \vert\, M'\in
\mathfrak v', \, k=1,2,\dots,n,\, \lambda\in \mathbb{R}\},
$$
is a complete commutative subset of $\mathbb{R}[\mathfrak v']^{SO(n')_{A'}}$, i.e,
\begin{equation}
\bar {L'}_{M'}^{\bar\Lambda'}=\bar L'_{M'}, \qquad \bar
L'_{M'}=\{\nabla_{M'} f\, \vert\, f\in\bar{\mathcal L}_{\mathfrak
v'}\}\subset \mathfrak j'_{M'}, \label{L3}
\end{equation}
for a generic $M' \in \mathfrak v'$. But from (\ref{mik}) and
(\ref{L3}) we also get that (\ref{L2}) holds for a generic $M' \in\mathfrak v'$.  $\Box$

\begin{remark} {\rm An alternative proof of theorem \ref{homogeneous} can be performed by using the compatibility
of Poisson brackets (\ref{*}) and (\ref{**}), but now considered within the algebra  $\mathbb{R}[\mathfrak v]^{SO(n)_A}$.}
\end{remark}

\subsection{Submersion of Manakov Flows} Let $\mathfrak A$ be
given by (\ref{A}) where $\mathfrak B$ is the identity operator.
Then the singular Manakov flow (\ref{s0}), (\ref{s1}) represent
the geodesic flow of the left $SO(n)$-invariant metric on $SO(n)$
that is also right $SO(n)_A$-invariant. By submersion, this metric
induces $SO(n)$-invariant metric on homogeneous space
$SO(n)/SO(n)_A$ that we shall denote by $ds^2_{A,B}$. Specially,
for $A=B$ we have the normal metric.

On $\mathfrak v$, identified with the tangent space at the class
of the identity element, the metric is given by the
$SO(n)_A$-invariant scalar product
\begin{equation}
( \,\cdot\,,\,\cdot \,)_{A,B}=\sum_{1\le i \le j \le r}
\frac{\alpha_i-\alpha_j}{\beta_i-\beta_j}\langle
\,\cdot\,,\,\cdot\, \rangle\vert_{\mathfrak v_{i,j}}, \label{EM1}
\end{equation}
where
$$
\mathfrak v=\bigoplus_{1\le i<j\le r} \mathfrak v_{i,j}
$$
is the decomposition into a sum of $SO(n)_A$-invariant subspaces
defined by $so(k_i+k_j)=so(k_i)\oplus so(k_j) \oplus \mathfrak
v_{i,j}$.

As before the formulation of theorem \ref{homogeneous},
let $\mathcal A$ be any commutative set of polynomial on
$\so(n)$ that is complete on adjoint orbits within the image of
the momentum mapping $\Phi$.
Since the Hamiltonian
$H_{A,B}(M)=\frac12\langle \ad^{-1}\circ\ad_B(M),M\rangle\in \mathbb{R}[\mathfrak v]^{SO(n)_A}\cong\mathcal F^{SO(n)}$
of the geodesic flow of the metric $ds^2_{A,B}$ Poisson commute with ${\mathcal L}_\mathfrak
v$, from theorem \ref{homogeneous} we get

\begin{corollary}
The geodesic flows of the metrics $ds^2_{A,B}$ on the homogeneous
spaces $SO(n)/SO(n)_A$ are completely integrable in the
noncommutative sense. The complete set of integrals is given by
(\ref{LV}) and Noether integrals $\Phi^*(\R[\so(n)])$. The
geodesic flows is also Liouville integrable by means of polynomial
integrals $\mathcal L_\mathfrak v+\Phi^*(\mathcal A)$.
\end{corollary}

\section{Examples: Einstein Metrics}

Among $SO(n)$-invariant metrics on $SO(n)/SO(k_1)\times\dots\times
SO(k_r)$ the specific geometric significance have Einstein metrics
(see \cite{Be}). It is well known that the unique (up to
homotheties) $SO(n)$-invariant metrics on symmetric spaces
$SO(n)/SO(n-k)\times SO(k)$ are Einstein. Further examples are
given by Jensen \cite{Jen}, Arvanitoyeorgos,  Dzhepko and
Nikonorov \cite{adn, ADN} and Nikonorov \cite{Nik} on Stiefel
manifolds $V(n,k)=SO(n)/SO(k)$ and spaces
$SO(k_1+k_2+k_3)/SO(k_1)\times SO(k_2)\times SO(k_3)$,
respectively.

The geodesic flows on symmetric spaces are completely integrable
(see Mishche\-nko \cite{Mis}). It is very interesting that the
geodesic flows of Einstein metrics given in \cite{Jen, Nik, adn,
ADN} are also integrable.

Firstly, note that the metrics on $SO(k_1+k_2+k_3)/SO(k_1)\times
SO(k_2)\times SO(k_3)$ constructed by Nikonorov in \cite{Nik} are
already of the form (\ref{EM1}). On the other side, to prove the
integrability of geodesic flows of Einstein metrics on Stiefel
manifolds $V(n,k)$ obtained in \cite{Jen, adn, ADN} we need the
following simple modification of theorem \ref{homogeneous}.

Let us fix $l$, $1 \le l \le r$ and consider products $SO(n)_A =H\times K$ and $\so(n)_A=\mathfrak h \oplus
\mathfrak k$, where
\begin{eqnarray*}
&&H=SO(k_1)\times\dots\times SO(k_l), \quad
K=SO(k_{l+1})\times\dots\times SO(k_r), \\
&& \mathfrak h=\so(k_1)\oplus\dots\oplus \so(k_l), \quad \qquad \mathfrak
k=\so(k_{l+1})\oplus\dots\oplus \so(k_r).
\end{eqnarray*}

Let
$
\mathfrak p=\mathfrak k \oplus \mathfrak v,
$
$\mathbb{R}[\mathfrak p]^{H}$ be the algebra of $\Ad_H$-invariant
functions on $\mathfrak p$ identified with the algebra of $SO(n)$
functions on $T^*(SO(n)/H)$, $\mathcal K$ be the algebra of linear
functions on $\mathfrak k$ lifted to the functions in
$\mathbb{R}[\mathfrak p]^{H}$ and
\begin{equation}
{\mathcal L}_\mathfrak p=\{\tr(M+\lambda A)^k\, \vert\, M\in
\mathfrak p, \, k=1,2,\dots,n,\, \lambda\in \mathbb{R}\},
\label{EM2}\end{equation}

\begin{theorem}
\label{homogeneous2} {\rm (i)} ${\mathcal L}_\mathfrak p+\mathcal
K$ is a complete subset of  $\mathbb{R}[\mathfrak p]^{H}$.

{\rm(ii)} If $\mathcal K^0$ is any complete commutative set of
functions on $\mathfrak k$ lifted to the functions on $\mathfrak
p$, then ${\mathcal L}_\mathfrak p+\mathcal K^0$ will be a
complete commutative subset of $\mathbb{R}[\mathfrak p]^{H}$.
\end{theorem}

Now, consider the case $r=2$, $l=1$, $H=SO(k)$, $K=SO(n-k)$. Then
$\mathfrak v_{1,2}=\mathfrak v$ and $\mathfrak
p=\so(n-k)\oplus\mathfrak v$. Define the $SO(n)$-invariant  metric
$ds^2_{\mathfrak I,\kappa}$ on $V(n,k)=SO(n)/SO(k)$ by its
restrictions to $\mathfrak p$:
\begin{equation}
(\,\cdot\,,\,\cdot\,)_{\mathfrak I,\kappa}=\langle
\,\cdot\,,\mathfrak I \,\cdot\,\rangle\vert_{\so(n-k)} + \kappa
\langle \cdot,\cdot \rangle \vert_{\mathfrak v}, \label{EM3}
\end{equation}
where $\mathfrak I: \so(n-k)\to \so(n-k)$ is positive definite and
$\kappa>0$.

Note that Manakov integrals (\ref{EM2}) are integrals of the
geodesic flow of the metric (\ref{EM3}). Thus, if Euler equations
on $\so(n-k)$
\begin{equation}
\dot M=[M,\mathfrak I^{-1} (M)]  \label{EM4}
\end{equation}
are integrable, the geodesic flow of the metric $ds^2_{\mathfrak
I,\kappa}$ will be completely integrable.

Let $\mathfrak I=\chi \cdot \mathrm{Id}_{\so(n-r)}$. In
\cite{Jen}, Jensen proved that for $n-k=2$ there is a unique
value, while for $n-k>2$ there are exactly two values of
$(\chi,\kappa)\in \R^2$ (up to homotheties), such that
$ds^2_{\mathfrak I,\kappa}$ is Einstein metric. Since then
equations (\ref{EM4}) are trivial, functions $\mathcal L_\mathfrak
p+\mathcal K$ are integrals of the geodesic flow.

Arvanitoyeorgos, Dzhepko and Nikonorov found two new Einstein
metrics \cite{adn, ADN} within the class of metrics (\ref{EM3})
with $n-k=sl$, $s>1$, $k>l \ge 3$. It appears that the
integrability of corresponding Euler equations (\ref{EM4}) can be
easily proved by using the chain method developed by Mykytyuk
\cite{Mik}.

\begin{corollary}
The geodesic flows of Einstein metrics on Stiefel manifolds
$SO(n)/$ $SO(k)$ and homogeneous spaces
$SO(k_1+k_2+k_3)/SO(k_1)\times SO(k_2)\times SO(k_3)$ constructed
in \cite{Jen, adn, ADN, Nik} are completely integrable.
\end{corollary}

Note that the integrability of the geodesic flows of Einstein
metrics on Stiefel manifolds $V(n,k)$ can be proved in a different
way, starting from the analogue of the Neumann system on $V(n,r)$
(see \cite{FeJo}).

\section*{Apendix: Pairs of Reductive Lie Algebras}

Let $\mathfrak g$ be a reductive real (or complex) Lie algebra. Take  a faithful
representation of $\mathfrak g$ such that its associated
bilinear form $\langle \cdot,\cdot \rangle$ is nondegenerate on $\mathfrak g$.
Let $\mathfrak k \subset \mathfrak g$ be a reductive in $\mathfrak g$ subalgebra and
$$
\mathfrak v=\mathfrak k^\perp= \{\eta\in \mathfrak g\, \vert\,
\langle \eta,\mathfrak k\rangle=0\}.
$$
For any $\xi\in \mathfrak v$ define the subspace $\mathfrak j_\xi \subset \mathfrak v$ by
$$
\mathfrak j_\xi=\{\eta\in\mathfrak v\,\vert\,[\xi,\eta]\in\mathfrak v\}=
\{\eta\in\mathfrak v\,\vert\, \langle \eta, [\xi,\mathfrak k]\rangle=0\}.
$$

Consider a Zariski open subset of $R$-elements in $\mathfrak v$ defined by
$$R(\mathfrak v)=\{\xi \in\mathfrak v\, \vert\, \dim \mathfrak g_\xi  \le \dim \mathfrak g_\eta,
\dim \mathfrak k_\xi  \le \dim \mathfrak k_\eta, \dim \mathfrak
g_\xi^0  \le \dim \mathfrak g_\eta^0, \, \eta\in\mathfrak v\},$$
where $\mathfrak g_\eta$ and $\mathfrak k_\eta$ are centralizers
of $\eta$ in $\mathfrak g$ and $\mathfrak k$, respectively, and
$\mathfrak g_\eta^0$ denote  the set of all $\zeta \in\mathfrak g$
which satisfy $(\ad\eta)^n(\zeta)=0$ for sufficiently large $n$.

Assume that $\xi_0\in R(\mathfrak v)$ and $\mathfrak a$ is a
reductive (in $\mathfrak g$) subalgebra of $\mathfrak k_{\xi_0}$.
Let $\mathfrak g'$ and $\mathfrak k'$ be the centralizers of
$\mathfrak a$ in $\mathfrak g$ and $\mathfrak k$, respectively.
Then algebras $\mathfrak g'$ and $\mathfrak k'$ are subalgebras
reductive in $\mathfrak g$ and the restriction of $\langle
\cdot,\cdot \rangle$ to $\mathfrak g'$ and $\mathfrak k'$ are
nondegenerate (for more details, see Mykytyuk \cite{M})

Let $\mathfrak v'$ be the orthogonal complement of $\mathfrak k'$ in $\mathfrak g'$.
Then $\mathfrak v'=\mathfrak g' \cap \mathfrak v$ \cite{M}. As above, define
$$
\mathfrak j'_\xi=\{\zeta\in\mathfrak v'\,\vert\,[\xi,\zeta]\in\mathfrak v'\}, \quad \xi\in \mathfrak v'
$$
and the set of $R$-elements $R(\mathfrak v')$ in $\mathfrak v'$.

The following result is contained in the proof of
theorem 11 \cite{M} (see also proposition 2.3 given in \cite{MP}).

\begin{theoremA} \label{MYK} {\rm(Mykytyuk~\cite{M})}
The relation
\begin{equation*}
\mathfrak j_\xi=\mathfrak j'_\xi
\label{MYK*}\end{equation*}
is satisfied for any element $\xi$ in a Zariski open subset $R(\mathfrak v')\cap R(\mathfrak v)$ of $\mathfrak v'$.
\end{theoremA}

\subsection*{Acknowledgments}
We are grateful to Alexey V. Bolsinov and Yuri G. Nikonorov on
useful discussions. We would also use the opportunity to thank the
referees for their remarks and comments which helped us to improve
the exposition. The results are presented at the conference
Transformation Groups, Moscow 2007. We would like to thank the
organizers for kind hospitality.

The research was supported by the Serbian Ministry of Science
Project 144014 Geometry and Topology of Manifolds and Integrable
Dynamical Systems.

\

\small

\sc

Vladimir Dragovi\' c

Mathematical Institute  SANU

 Kneza Mihaila 36, 11000 Belgrade, Serbia

\& GFM, University of Lisbon, Portugal

{\rm e-mail: vladad@mi.sanu.ac.rs}

\

Borislav Gaji\' c

Mathematical Institute  SANU

Kneza Mihaila 36, 11000 Belgrade, Serbia

{\rm e-mail: gajab@mi.sanu.ac.rs}

\

Bo\v zidar Jovanovi\'c

Mathematical Institute  SANU

 Kneza Mihaila 36, 11000 Belgrade, Serbia

{\rm e-mail: bozaj@mi.sanu.ac.rs}

\end{document}